\documentclass[%
 reprint,
 superscriptaddress,
 amsmath,amssymb,
 aps,
]{revtex4-2}
\usepackage[nodisplayskipstretch]{setspace} 
\usepackage{graphicx}
\usepackage{dcolumn}
\usepackage{bm}
\usepackage{braket}
\usepackage{color}
\usepackage{hyperref} 

\usepackage{fancyhdr}
\begin{document}
\fancyhead[R]{\ifnum\value{page}<2\relax\else\thepage\fi}

\preprint{APS/123-QED}

\title{Physics Simulation Via Quantum Graph Neural Network}

\author{Benjamin Collis}
\email{bjmcollis@gmail.com}
\affiliation{Griffiss Institute, Information Directorate, Rome, NY}
\affiliation{Air Force Research Laboratory, Information Directorate, Rome, NY}

\author{Saahil Patel}
\affiliation{Air Force Research Laboratory, Information Directorate, Rome, NY}

\author{Daniel Koch}
\affiliation{Air Force Research Laboratory, Information Directorate, Rome, NY}

\author{Massimiliano Cutugno}
\affiliation{Air Force Research Laboratory, Information Directorate, Rome, NY}

\author{Laura Wessing}
\affiliation{Air Force Research Laboratory, Information Directorate, Rome, NY}

\author{Paul M. Alsing}
\affiliation{Air Force Research Laboratory, Information Directorate, Rome, NY}

%
%
\begin{abstract}
We develop and implement two realizations of quantum graph neural networks (QGNN), applied to the task of particle interaction simulation. The first QGNN is a speculative quantum-classical hybrid learning model that relies on the ability to directly utilize superposition states as classical information to propagate information between particles. The second is an implementable quantum-classical hybrid learning model that propagates particle information directly through the parameters of $RX$ rotation gates. A classical graph neural network (CGNN) is also trained in the same task. Both the Speculative QGNN and CGNN act as controls against the Implementable QGNN. Comparison between classical and quantum models is based on the loss value and accuracy of each model. Overall, each model had a high learning efficiency, in which the loss value rapidly approached zero during training; however, each model was moderately inaccurate. Comparing performances, our results show that the Implementable QGNN has a potential advantage over the CGNN. Additionally, we show that a slight alteration in hyperparameters in the CGNN notably improves accuracy, suggesting that further fine tuning could mitigate the issue of moderate inaccuracy in each model.

\end{abstract}

\maketitle

\thispagestyle{fancy}

\section{\label{sec:level1}Introduction}

Irregular graph-based machine learning is inherently difficult due to the lack of symmetry among the nodes and edges that constitute the graph \cite{Hamilton}. Convolutional neural networks (CNN) that thrive in the context of grid-based data are ineffective and inappropriate to use in this context, as they lack a straight-forward method of incorporating irregular data \cite{Hamilton}. This is exemplified in cases where a single node may have hundreds to thousands of edges, whereas its neighbor only has one. Recent pursuit of a neural network model that adapts to the complexities implicit in irregular graph-based systems, e.g. social networks \cite{Fan}, molecule structures \cite{Duvenaud}, particle interactions \cite{Battaglia_Interaction}, etc., has resulted in the development of a plethora of graph-based machine learning models which are all defined under the general term ``graph neural network" (GNN) \cite{Battaglia_Relational}. Though containing a spectrum of deviations, the core feature of these models is that they exchange information between nodes via vector messages, a process dubbed message passing, and then update via neural network \cite{Hamilton}. It is this core feature that has lead to the success of GNNs, as evidenced by their ability to excel at a variety of tasks including predictions ranging from traffic \cite{Cui}, to the chemical properties of molecules \cite{Duvenaud}, knowledge graph reasoning \cite{Bordes} and particle simulation \cite{Battaglia_Interaction}. In essence, the utility of GNNs encompasses a majority of learning problems that can be represented as a graph containing meaningful connections between nodes. The vagueness here is appropriate due to the extensive number of environments where GNNs are applicable \cite{Hamilton, Battaglia_Relational}.

Quantum machine learning has likewise emerged recently with developments in quantum hardware sophistication and capacity \cite{Biamonte}, and is itself a sub-field of machine learning that has garnered increased interest over the last decade. Being relatively new and having an excess of possible realizations, there is a lack of formal definition for the topic \cite{Schuld}. In general, the process involves traditional learning via quantum information processing, where either parameters are optimized or a decision function is obtained \cite{Schuld}. For the purposes of this study, the quantum machine learning aspects are reserved to encoding, processing, and decoding the data via parameterized quantum circuits, while the remaining parts of the algorithms including calculating loss function, back-propagation, etc., are done classically. Thus, the learning models presented here represent hybrid quantum-classical algorithms.
\begin{figure*}[!th]
\includegraphics[width=\textwidth,height=\textheight,keepaspectratio]{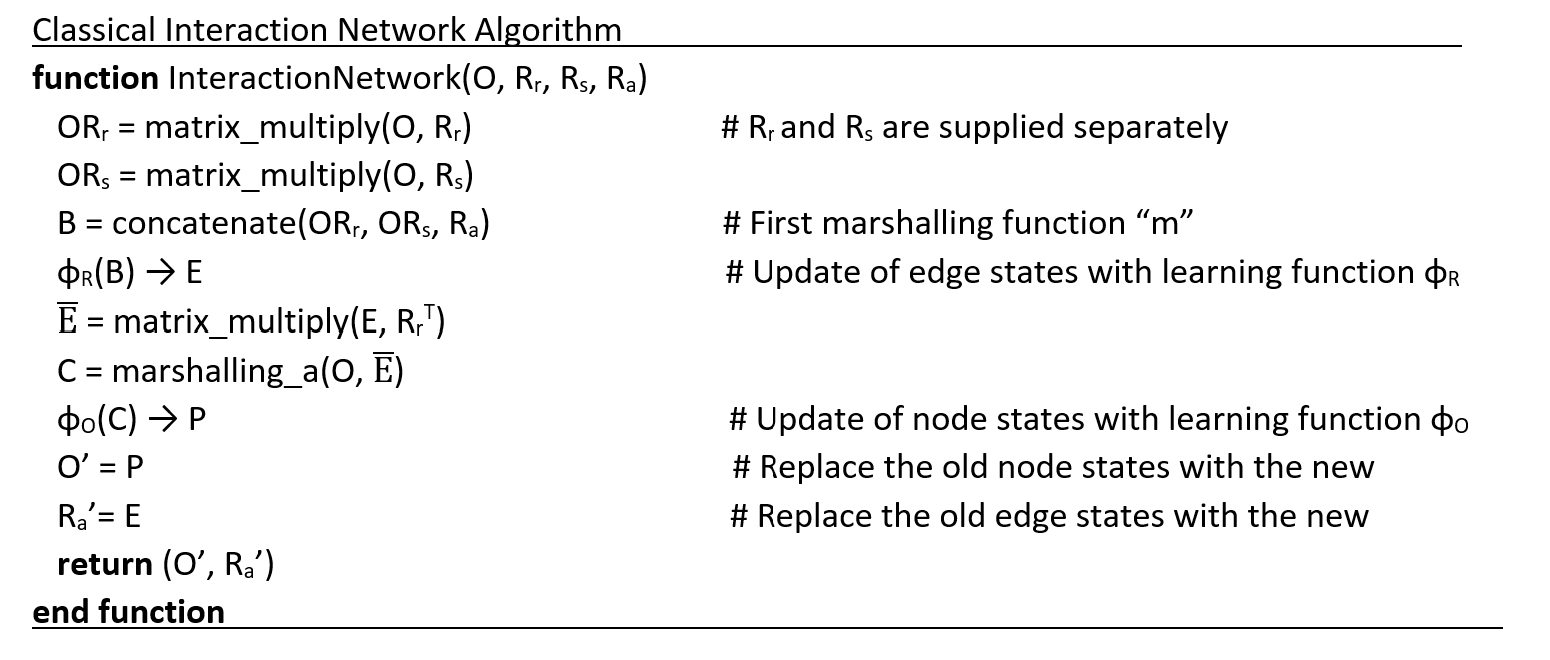}
\caption{Pseudocode of the classical interaction network algorithm}
\centering
\label{fig:classical_algo}
\end{figure*}

A general pursuit in quantum algorithms is quantum supremacy \cite{Preskill}. Yet prior to even attempts at such a lofty claim, it must be ascertained as to whether a quantum analog is obtainable for a classical algorithm. That is the main pursuit of this study; following confirmation of this, we consider performance comparisons between the classical and quantum cases. We begin by describing and implementing two quantum graph neural network (QGNN) learning models. In particular, the QGNN consists of three sections of parameterized quantum circuits (PQC): encoder, processor, and decoder. The encoder expands the initial superposition state by inducting additional qubits into it, and the decoder pools that information from the larger number of qubits into the desired smaller amount (pooling from six qubits to two). The processor in between is responsible for the message passing, utilizing a quantum-based interaction network (IN) to send information between nodes \cite{Battaglia_Interaction}. The variation in the two developed QGNNs is that the first is a speculative model, while the second is implementable. Specifically, the speculative model relies on being able to directly take and store the qubits' superposition amplitude states and to use them classically. This is not possible to directly implement on a quantum computer, as a superposition state can only be approximated via statistical analysis of numerous measurements \cite{Lvovsky}. For the speculative circuit, the statistical analysis would have to be implemented seven times for each input vector, where seven is the number of sub-circuits that constitute the overall quantum circuit, excluding the decoder. Supposing $1,000$ measurements per approximation to reconstruct the superposition state, the speculative circuit would need to be run $7,000$ times per input vector. Considering the number of input vectors for this study is approximately $19,000$, this would require $133$ million runs of the quantum computer for a single epoch. Furthermore, each run would require use of a subroutine to determine the sign of each superposition state's amplitude \cite{Lvovsky, Kerenidis}, in addition to needing a subroutine to generate the expectation value that constitutes the quantum circuit's relevant output \cite{Paini}. This is all very impractical for actual implementation on quantum hardware; thus, this model is dubbed speculative. Regardless, this quantum algorithm was designed to be the closest analog to the classical, and thus acts as a control. The Implementable QGNN is fully realizable on quantum hardware. However, due to considerable gate depth, the results shown in this study were achieved via a quantum circuit simulator. Both the Speculative and Implementable QGNNs, which they will be referred to as throughout this paper, correlate the output of a particular qubit to its expectation value following application of a Pauli-Z observable to that particular qubit \cite{pennylane_measurement, pennylane_measurement_2}. Actual use of the Implementable QGNN on a quantum computer requires a non-trivial subroutine to approximate the observable's expectation value \cite{Paini}. However, as this is only necessary at the end of the circuit and only requires application for two qubits in this study, it is considered implementable. The utility of these learning models is examined under the context of particle interaction simulation. In particular, the case of point particles falling under the influence of gravity within a box.

This study contains the following layout. Section II covers the learning algorithms. Section III covers the PQCs of the encoder, processor, and decoder. Additionally, it goes over the method of encoding the data, and it describes the progression of data from input to output for both QGNNs. Section IV covers the results of the QGNNs and CGNN. Section V concludes this study by considering the results and offering potential paths for future research.

\section{Learning Model}
\subsection{Overview}
The learning model implemented in this study is based on that used by Sanchez-Gonzalez et al.  \cite{Sanchez-Gonzalez}. It includes three sections: the encoder, the processor, and the decoder. The encoder takes the initial vector input and expands it into a higher dimensional latent space. The processor then processes the expanded data through its interaction network (IN) for a select $n$ number of steps. Each step corresponds to a particular node receiving a ``message" from nodes $n$ edges away. Lastly, the decoder receives this processed data and outputs a prediction \cite{Sanchez-Gonzalez}. 

\subsection{Classical Interaction Network}
The message passing property of the classical graph neural network (GNN) used in this project is obtained through use of the IN learning model. It is the same as that described by Battaglia et al. \cite{Battaglia_Interaction}, and is highly similar to the Graph Network (GN) learning model described by Sanchez-Gonzalez et al. \cite{Sanchez-Gonzalez}. A brief description of the classical IN is provided here, while a more in-depth analysis can be found in Battaglia et al.'s work \cite{Battaglia_Interaction}. Following this are the descriptions of the GNN quantum analogs, the quantum graph neural networks (QGNN). Appropriately, each QGNN has a unique quantum interaction network, each analogous to the classical.

As described by Battaglia et al. \cite{Battaglia_Interaction}, the classical IN is given below.

\begin{align}
    & IN_{Classical} = \phi_{O}(a(G,\phi_{R}(m(G)))) \label{phi_o}
\end{align}

The IN contains two neural networks, $\phi_{O}$ and $\phi_{R}$. Respectively, these represent the node and edge state updates functions. The other two functions are the marshalling functions $a$ and $m(G)$, which concatenate the data supplied to them \cite{Battaglia_Interaction}. For $m(G)$, this data consists of $G$, which is described as follows:

\begin{align}
    & G = \braket{O,R} = \notag\\
    & \braket{\{o_{j}\}_{j=1...N_{O}},\braket{\{R_{r}\}_{k},\{R_{s}\}_{k},\{R_{a}\}_{k}}_{k=1...N_{R}}}, \label{class_G}
\end{align}

where $O$ is the set of $N_{O}$ nodes in the graph with state vector length $O_{l}$, and $R$ is the set of $N_{R}$ directed edges with state vector length $R_{l}$. Thus, $O$ is a matrix of size $O_{l} \times N_{O}$. The set $R$ can be further decomposed into the triple set $R_{r}$, $R_{s}$, $R_{a}$, where for a given pair of nodes connected by a directed edge, $R_{r}$ is the receiver node, $R_{s}$ is the sender node, and $R_{a}$ is the state of that edge. $R_{r}$, $R_{s}$, and $R_{a}$ are the respective matrix representations that contain all the receiver, sender, and edge state information of the graph. $R_{r}$ and $R_{s}$ contain only $0$s and $1$s, and are each of size $N_{O} \times N_{R}$, where row index $j$ corresponds to node $o_{j}$, and column index $k$ corresponds to edge $\{R_{a}\}_{k}$. For $R_{r}$, a $1$ corresponds to a node being the receiver of a particular edge. Likewise, for $R_{s}$, a $1$ corresponds to a node being the sender of a particular edge. $R_{a}$ is a matrix of size $R_{l} \times N_{R}$, where the columns are the edges with state vector length $R_{l}$. The state vector length is arbitrary for both edges and nodes. Thus, $G$ defines the state of a graph, containing complete information of the nodes and their connections.

The output $B$ of the marshalling function $m(G)$ is described below, along with its column slices, $b_{k}$.
\begin{align}
    & m(G) = conc[OR_{r}; OR_{s}; R_{a}] = B\hspace{1mm};\hspace{1mm} b_{k} \subset B \label{b_sub}\\
    & \phi_{O}(a(G,\phi_{R}(m(G)))) =  \phi_{O}(a(G,\phi_{R}(B)))
\end{align}
$B$ consists of the concatenation of the matrix multiplications of $OR_{r}$ and $OR_{s}$, combined with $R_{a}$ \cite{Battaglia_Interaction}. This packages the node and edge information in a convenient way for implementation into the neural network. In particular, $B$ is a matrix composed of $OR_{r}$ stacked on top of $OR_{s}$ stacked on top of $R_{a}$. With $OR_{r}$ and $OR_{s}$ both taking the shape $O_{l} \times N_{R}$, and $R_{a}$ taking the shape $R_{l} \times N_{R}$, the combination results in the $B$ matrix taking the shape $(2O_{l} + R_{l}) \times N_{R}$.
\begin{figure*}[!th]
\includegraphics[width=\textwidth,height=\textheight,keepaspectratio]{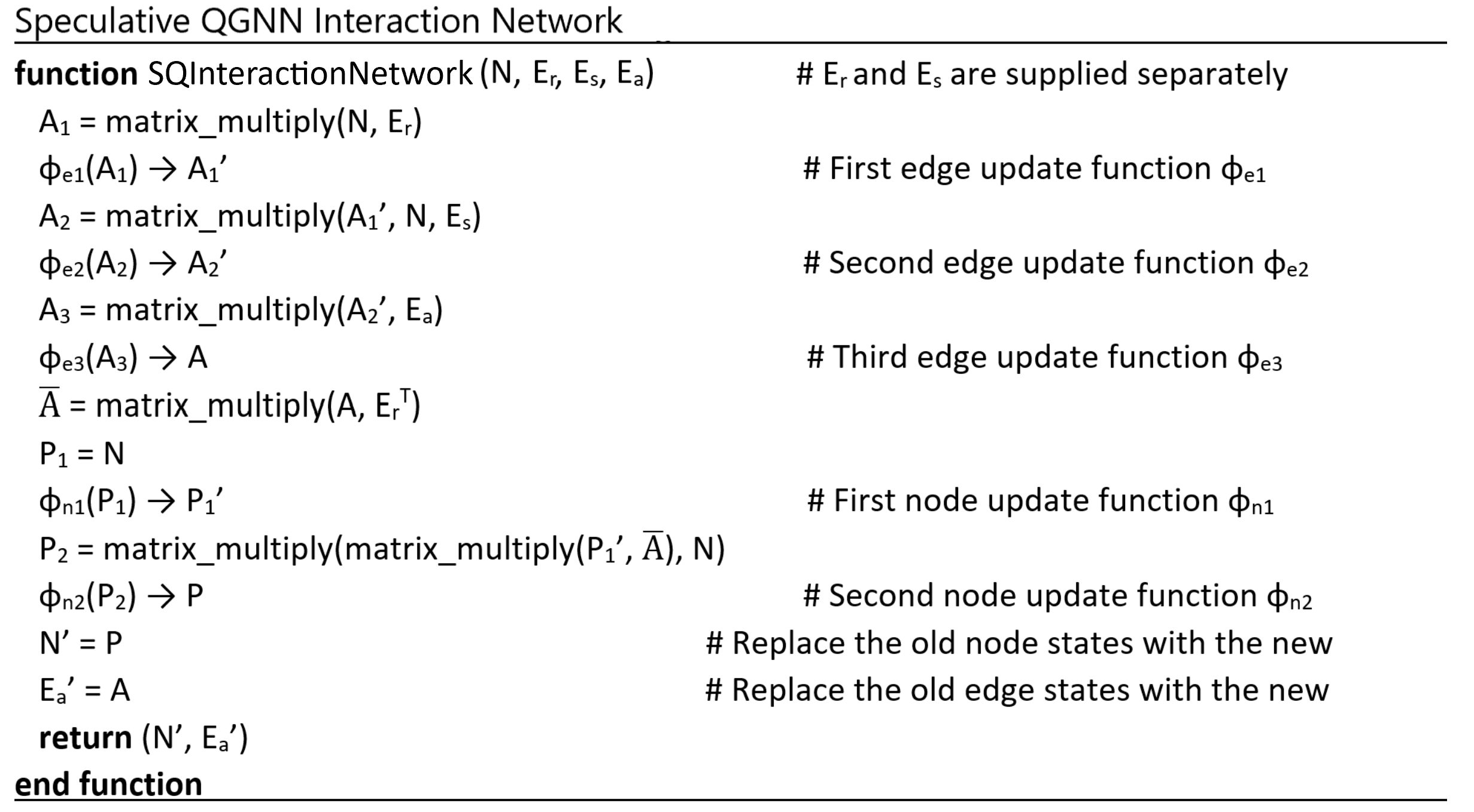}
\caption{Pseudocode of the Speculative QGNN interaction network algorithm}
\centering
\label{fig:YES_QRAM_algo}
\end{figure*}
\begin{align}
    & \phi_{R}(B) = E \\
    & \phi_{O}(a(G,\phi_{R}(B))) =  \phi_{O}(a(G,E))
\end{align}

Next, $B$ is supplied to $\phi_{R}$. Its column slices, $b_{k}$ (see equation \ref{b_sub}), are the input, and the total number of columns in $B$ is the batch size. $\phi_{R}$ predicts the new edge states $E$, which being a matrix of the edges containing only new edge states, is the same size as $R_{a}$. This is appropriate, as $E$ will replace $R_{a}$ in the case of multiple processors, i.e. multiple iterations of this learning algorithm. 
\begin{align}
    & \bar{E} = ER^{T}_{r} \\
    & \phi_{O}(a(G,E)) = \phi_{O}(a(O,\bar{E})) \label{equiv}
\end{align}

The transformation of $E$ into $\bar{E}$, with shape $N_{R} \times N_{O}$, is useful in that it combines the information of the new edges and the nodes that this update will effect \cite{Battaglia_Interaction}. Furthermore, it allows for the equivalency in equation \ref{equiv}, such that $(O,\bar{E})$ contains the same relevant information including node states, edge updates, and nodes impacted as found within $(G,E)$.
\begin{align}
    & a(O,\bar{E}) = C ; c_{k} \subset C \label{c_sub}
\end{align}

The second marshalling function, $a$, concatenates the columns of $O$ and $\bar{E}$, outputting matrix $C$. For the $k$-th column of $C$, the $k$-th column of $O$ is the top half and the $k$-th column of $\bar{E}$ is the bottom half. In effect, $C$ is composed of $O$ on top of $\bar{E}$. This packages the node and edge information in a convenient way for implementation into the neural network \cite{Battaglia_Interaction}. Again, $O$ has shape $O_{l} \times N_{O}$ and $\bar{E}$ has shape $N_{R} \times N_{O}$; thus, the size of matrix $C$ is $(O_{l} + N_{R}) \times N_{O}$.
\begin{align}
    & \phi_{O}(a(O,\bar{E})) = \phi_{O}(C) = P
\end{align}

$C$ is then supplied to the learning function $\phi_{O}$. Its columns' slices, $c_{k}$ (see equation \ref{c_sub}), are the input, and the total number of columns in $C$ is the batch size. Just as $\phi_{R}$ predicted the new edge states, $\phi_{O}$ now predicts the new node states, $P$, which is a matrix the same size as $O$ that contains the new node states \cite{Battaglia_Interaction}. This is appropriate, as $P$ will replace $O$ in the case of multiple processors. Finally, either the algorithm repeats, or the matrix $P$ is supplied to the decoder. The outcome is dependent upon the number of processors in the GNN, with each processor corresponding to one complete run of the IN. For the case of multiple processors, the substitution $O' = P$ and $R'_{a} = E$ will be made, and the algorithm will be repeated with these updated $O$ and $R_{a}$ values. After cycling through all processors, the final $P$ is given to the decoder, which will then output its prediction. The psuedocode for the full classical IN algorithm is shown in figure \ref{fig:classical_algo}. Additionally, a complete explanation of this algorithm can be found in the work of Battaglia et al. \cite{Battaglia_Interaction}.

To give a brief overview of the classical GNN implemented in this study, it has the overall same design as that utilized by Sanchez-Gonzalez et al. \cite{Sanchez-Gonzalez}., and is described as follows. The node encoder consists of a two layer multilayer perceptron (MLP), with the first layer being size $8$ and the second layer being size $9$. The edge encoder also consists of a two layer MLP, with the first layer being size $4$ and the second layer being size $5$. The first layers for these sections were explicitly chosen to correspond to the node and edge input data, which are of size $8$ and $4$, respectively. Concerning the processors, the node and edge processor each consist of a single perceptron layer of size $9$ and $5$, respectively. Last, the decoder contains a single perceptron layer of size $2$. After each perceptron layer in the encoders and processors, layer normalization is used \cite{Jimmy}. Additionally, each of these layers utilize a ReLU activation function, while the decoder does not implement one.

\subsection{Speculative QGNN Interaction Network}
The IN algorithm implemented in the Speculative QGNN behaves as follows:
\begin{align}
    & IN_{Speculative} = \phi_{n}(S_{n}(G,\phi_{e}(S_{e}))) \label{phi_n}
\end{align}

The algorithm consists of two learning functions, $\phi_{n}$ and $\phi_{e}$, with the latter embedded within the prior. The learning function $\phi_{e}$ is responsible for updating the edge states, i.e. predicting the effects of nodal connections. Likewise, $\phi_{n}$ is the learning function responsible for updating the node states. The graph G, which is the same as the classical case, is described again as
\begin{align}
    & G = \braket{N,E} = \notag\\
    & \braket{\{n_{j}\}_{j=1...N_{n}},\braket{\{e_{r}\}_{k},\{e_{s}\}_{k},\{e_{a}\}_{k}}_{k=1...N_{e}}}, \label{G}
\end{align}
\begin{figure*}[!th]
\includegraphics[width=\textwidth,height=\textheight,keepaspectratio]{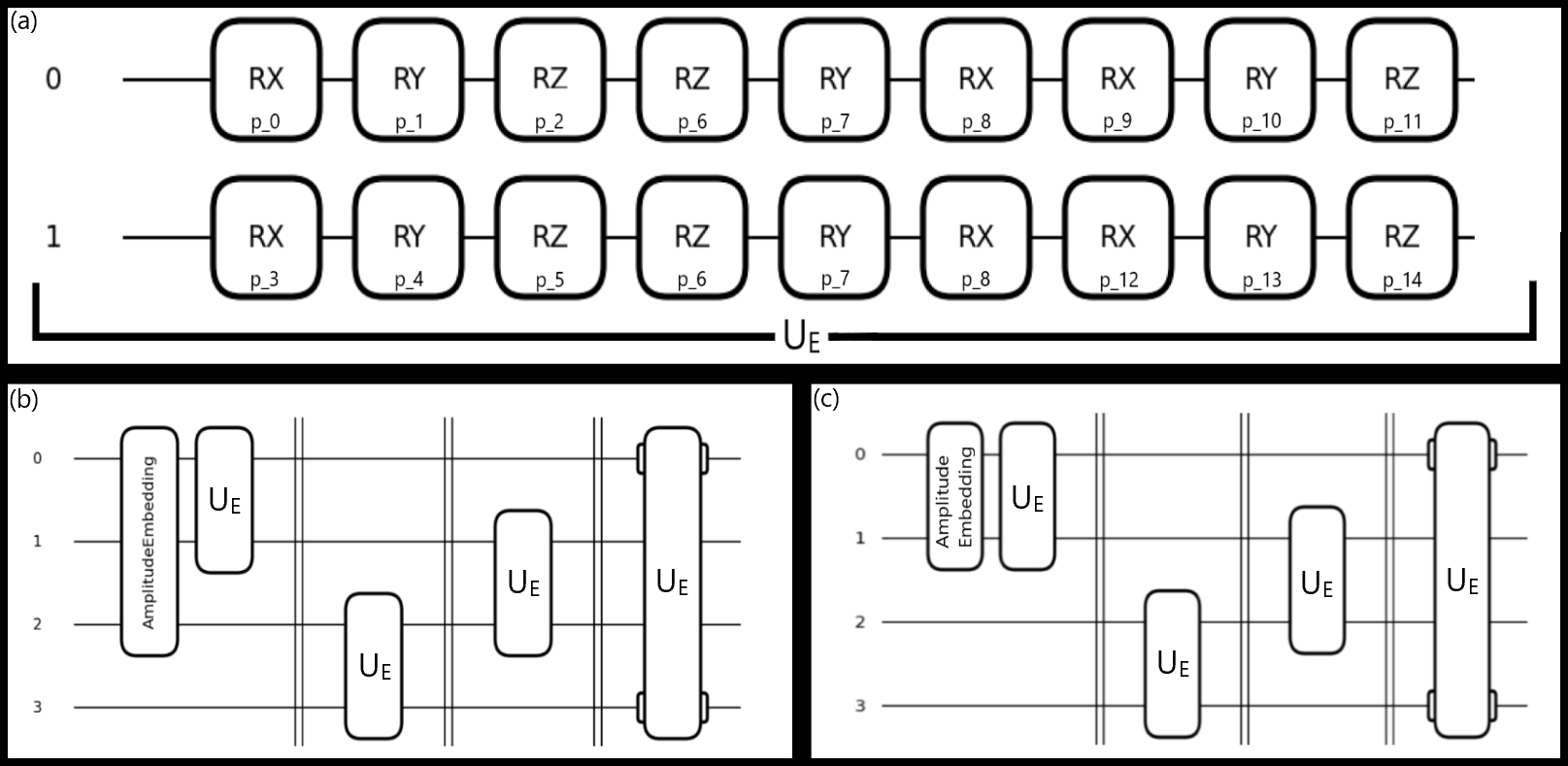}
\caption{(a) The composition of the encoder convolution unitary. (b) The application of the encoder convolution unitary in encoding the node state input. (c) The application of the encoder convolution unitary in encoding the edge state input. Note, the last encoder unitary applied in (b) and (c) indicates that the top half of gates in (a) is applied to qubit $3$ and the remaining bottom half is applied to qubit $0$, i.e., it is applied in the same exact manner as the prior encoder unitaries.}
\centering
\label{fig:encoder}
\end{figure*}

Note the simple change in variable names for the quantum case as opposed to the classical. In particular, $O$ is equal to $N$ and $R$ is equal to $E$. This algorithm contains steps of matrix multiplication, which is where the superposition states are directly implemented. In particular, the outputs of the learning functions at each time step are the superposition states, which are used to construct their corresponding matrices. These are then used to propagate information via matrix multiplication. 

\begin{align} 
    & A_{1} = NE_{r}\hspace{1mm}  ;\hspace{1mm}  A_{2} = 
    A_{1}^{'}NE_{s}\hspace{1mm}  ;\hspace{1mm}  A_{3} = A_{2}^{'}E_{a} \label{A_n}\\
    & A_{1}^{'} = \phi_{e_{1}}(A_{1}) \hspace{1mm}  ;\hspace{1mm} A_{2}^{'} = \phi_{e_{2}}(A_{2}) \label{A_n'}\\
    & S_{e} = \braket{\phi_{e_{1}}(A_{1}),\phi_{e_{2}}(A_{2}), \phi_{e_{3}}(A_{3})} \label{S_E}\\
    & \phi_{e}(S_{e}) = \phi_{e_{1}}(A_{1}) \rightarrow  \phi_{e_{2}}(A_{2})  \rightarrow  \phi_{e_{3}}(A_{3}) = A \label{phi_S_e}
\end{align}

Classically, $S_{e}$ would be a marshalling function that concatenates $A_{1}$, $A_{2}$, and $A_{3}$, shown in equations \ref{A_n} and \ref{A_n'}, into a vector of size $(2N_{l} + E_{l}) \times N_{e}$. However, the quantum circuits do not adjust well to spontaneous changes in vector size, with quantum data compression of the superposition state being a non-trivial task \cite{Romero,Saahil}. Thus, it was optimal to alter $S_{e}$ to represent the application of $A_{1}$, $A_{2}$, and $A_{3}$ in series to $\phi_{e}$, which has been decomposed into three separate learning functions $\phi_{e1}$, $\phi_{e2}$, and $\phi_{e3}$. The output of this series of learning functions is $A$, the matrix of updated edge states, which is equivalent to the overall output of $\phi_{e}$.
\begin{align} 
    & \bar{A} = AE_{r}^T \label{A_BAR} \\
    & \phi_{n}(S_{n}(G,A)) \rightarrow \phi_{n}(S_{n}(N,\bar{A})) \label{phi_n}
\end{align}

$(G, A)$ describes the nodal composition of the graph, including the corresponding directed edges and their predicted effects. With the transformation of $A$ to $\bar{A}$, $(N, \bar{A})$ contains the same information. Thus, the substitution $(G, A)$ $\rightarrow$ $(N, \bar{A})$ is a convenient method of sorting the data for implementation.
\begin{align}
    & P_{1} = N \hspace{1mm};\hspace{1mm} P_{2} = (P_{1}^{'}\bar{A})N \label{P_1}\\
    & P_{1}^{'} = \phi_{n1}(P_{1})\\
    & S_{n} = \braket{\phi_{n_{1}}(P_{1}), \phi_{n_{2}}(P_{2})} \label{S_N} \\ 
    & \phi_{n}(S_{n}) = \phi_{n_{1}}(P_{1}) \rightarrow \phi_{n_{2}}(P_{2}) = P \label{phi_S_n}
\end{align}

$S_{n}$ performs the same process as $S_{e} $, except in the context of the nodal learning function where $S_{n}$ applies $P_{1}$ and $P_{2}$ in series to $\phi_{n}$, which has been decomposed into $\phi_{n1}$ and $\phi_{n2}$. The output of this series of learning functions is $P$, the updated node states, which can be designated the output of $\phi_{n}$.
\begin{align}
    & N' = P \label{N'}\\
    & E'_{a} = A \label{E'}
\end{align}
\begin{figure*}[!th]
\includegraphics[width=\textwidth,height=\textheight,keepaspectratio]{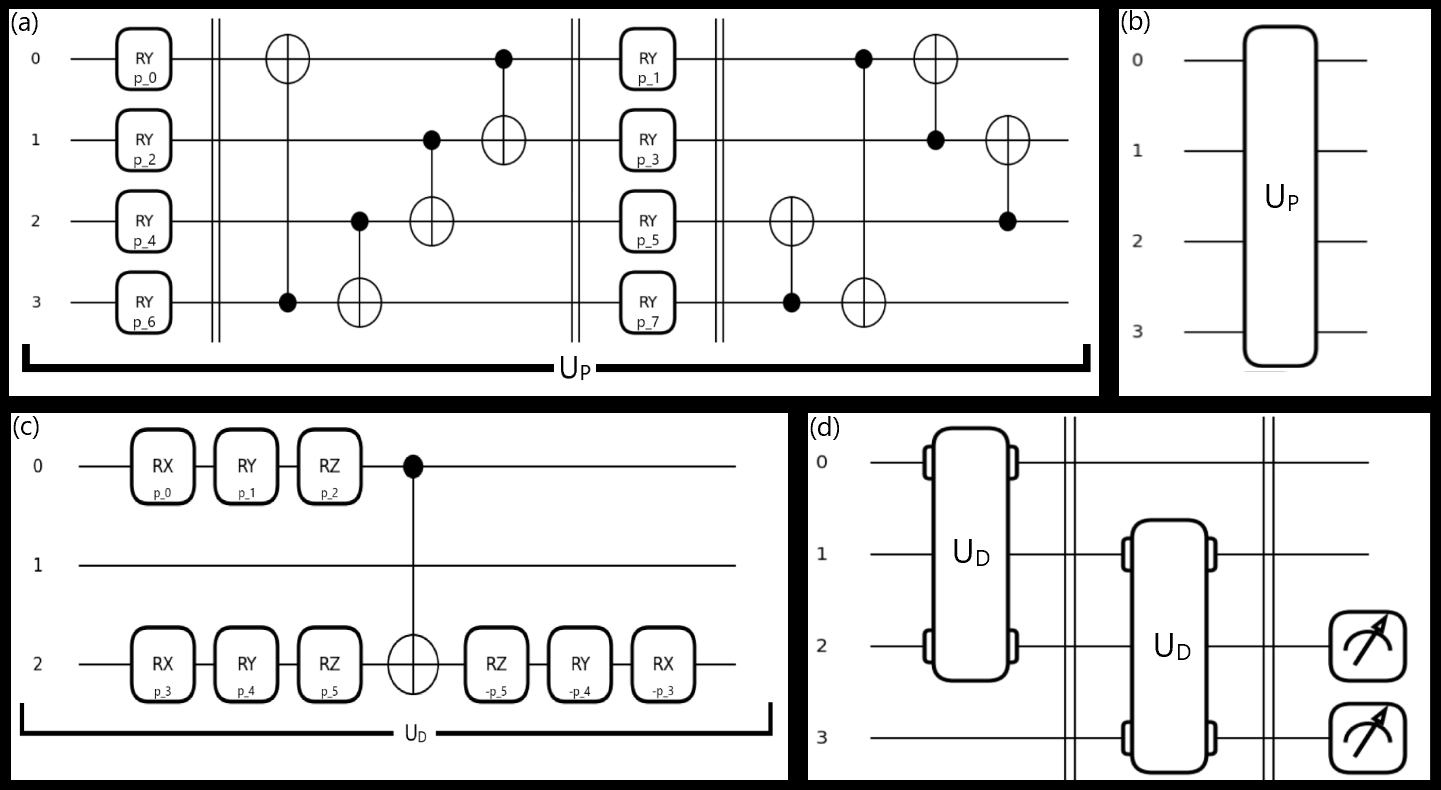}
\caption{(a) The composition of the processor PQC and (b) its corresponding unitary representation. (c) The composition of the decoder PQC and (d) its corresponding unitary representation}
\centering
\label{fig:processor/decoder}
\end{figure*}
The next step is to either rerun the algorithm with the updated node and edge states, i.e. equations \ref{N'} and \ref{E'}, or have the update features proceed to the decoder. This decision is based on the chosen number of runs, and the particular run the algorithm is on. For additional insight, figure \ref{fig:YES_QRAM_algo} contains the psuedocode for the Speculative QGNN IN algorithm.

It should be noted that for the purposes of this project only the updated node states were input into the decoder, while the updated edge states were only used in the node state update function and, thus, were confined to use within this algorithm, i.e. the processor. This is based on the similar process followed by Sanchez-Gonzalez et al. \cite{Sanchez-Gonzalez}. Additionally in static graphs, the matrices $E_{r}$ and $E_{s}$ of the receiver and sender indices remain the same. However, this project relies on dynamic graphs, meaning $E_{r}$ and $E_{s}$ change with the time progression of the particle interactions. This progression is described by time steps, each one corresponding to the overall state of the system at a particular moment in time, represented by a graph. 

Thus, edges are uniquely constructed between nodes for each graph, which is achieved via a nearest neighbour algorithm within a particular ``connectivity" radius used for each node at each time step, as implemented by Sanchez-Gonzalez et al. \cite{Sanchez-Gonzalez}.

\begin{figure*}[!th]
\includegraphics[width=\textwidth,height=\textheight,keepaspectratio]{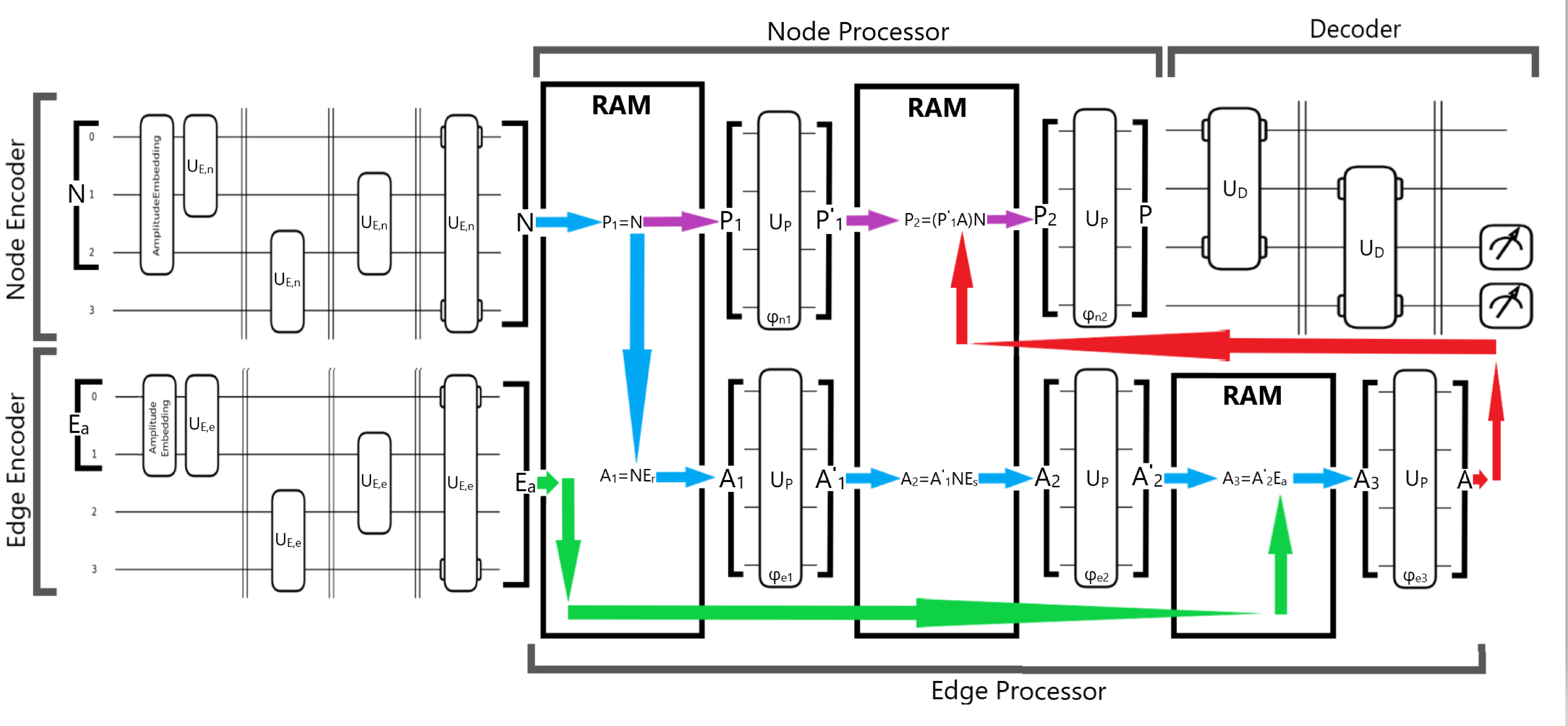}
\caption{\textbf{Speculative QGNN:} Depiction of the entire quantum circuit as it runs through the algorithm. The random access memory (RAM) sections represent where the superposition states are being saved and then used classically.}
\centering
\label{fig:full_PQC_and_ALGO}
\end{figure*}

\subsection{Implementable QGNN Interaction Network}
The design of the Implementable QGNN requires further deviation from the initial classical GNN learning model. In particular, this is a result of only being able to utilize slices of larger matrices during a single run-through of the entire algorithm. This is required in order to maintain superposition in the quantum circuit. Whereas the classical GNN and the Speculative QGNN both rely on two marshalling functions, the Implementable QGNN has none. Instead, it applies only a single column of $N$, $E_{r}$, $E_{s}$, and $E_{a}$ at a time, via a series of $RX$ gates to the qubits corresponding to the edges. The resultant information in the edge state qubits is then transitioned to into the node state qubits via decoding unitaries. Overall, this results in a method of information propagation not based in direct matrix multiplication, but instead in the application of rotation matrices. Furthermore, it adds the requirement of an additional layer of decoding unitaries.
Note that all of this does not suggest that the Implementable QGNN will necessarily be less effective, nor that it is somehow a worse implementation of GNNs. Instead, these dissimilarities from the Sanchez-Gonzalez et al. GNN \cite{Sanchez-Gonzalez} make the QGNN model studied here a novel attempt at quantum machine learning.

There are multiple steps in the Speculative QGNN learning model, i.e., a series of matrix multiplications. However, the Implementable QGNN learning model is more aptly defined by its unique design, in that there are no steps implemented outside the context of the quantum circuit. Thus, the entire algorithm is realized in a single quantum circuit, and is best explained through examination of the quantum gates that constitute it. The full description of the Implementable QGNN can be found in the Methods sections below. 

\section{\label{sec:level1}Methods}
\subsection{Data Encoding}

Each learning model is trained on two data sets, derived from the same classical simulation (see section IV.A), to create the initial node and edge state vectors. The initial state vector for a particular node consists of its previous two velocities, $v_{n-1}$ and $v_{n-2}$, and its normalized clipped distances to the boundaries, $b_{i}$, where these distances are clipped by the connectivity radius \cite{Sanchez-Gonzalez}. Each velocity has vector length two, and the vector sum of the clipped distances is length four. Concatenating these features gives the initial node state vector its size of eight, i.e., $(v_{x,n-1}, v_{y,n-1}, v_{x,n-2}, v_{y,n-2}, b_{1}, b_{2}, b_{3}, b_{4})$. A vector length of $8$ was chosen because values of $2^n$ are naturally easy to work with in a quantum circuit. Likewise, it is ideal to use a small number of qubits to avoid substantial training times and noise. The initial state vector for a particular edge consists of its relative positional displacement, $d_{x}$ and $d_{y}$, i.e. the distance between the corresponding sender and receiver given a particular edge, and that distance's corresponding magnitude, $D$; the prior is vector length two, and the latter is vector length one. Requiring an input size of $2^n$, a single layer of zero padding was added to each edge state vector. Concatenating these features gives the initial edge vector its size of four, i.e., $(d_{x}, d_{y}, D, 0)$. Outside of the zero padding, this composition of data is the same as that used by Sanchez-Gonzalez et al. \cite{Sanchez-Gonzalez}.

An initial step in quantum machine learning is encoding classical data into qubits, which can be accomplished using various methods such as qubit encoding \cite{Benedetti}, tensor product encoding, and amplitude encoding \cite{Roy}. The latter was implemented using the AmplitudeEmbedding function available in Pennylane, the quantum-compatible python package used to realize this project's classical-quantum algorithms \cite{pennylane_ae}. Amplitude encoding consists of embedding classical input values into the amplitudes of a quantum state. The requires transforming the data from its classical format into that of a superposition state. A superposition state consists of $2^n$ values, $n$ being the number of qubits for a given quantum circuit. Thus, the criterion arises for the input data to be of size $2^n$.

\subsection{Parameterized Quantum Circuits}

%
%

The Parameterized Quantum Circuits (PQC) used for the encoder and decoder are the same as those utilized in the Quantum Convolution Neural Network (QCNN) designed by Cong et al. \cite{Cong}, while the processor is the same as Circuit $15$ designed by Hubregtsen et. al \cite{Hubregtsen}. The QCNN PQCs are valuable in that they provide both a method for expanding data into a higher dimensional latent space and for pooling information into a desired number of qubits. Conversely, the value of Circuit $15$ is more holistic, being a PQC that was proven to be moderately accurate in the context of classification, while retaining a low number of required parameters.

The encoder is the reverse QCNN used by Cong et al., which is equivalent to the multiscale entanglement renormalization ansatz (MERA) \cite{Cong}. Thus, instead of pooling the information, it expands it to a higher dimensional latent space. This reverse QCNN is realized via the repeated application of a two qubit unitary, as shown in figure \ref{fig:encoder}a \cite{QCNN}. This unitary consists of $RX$, $RY$, and $RZ$ gates, with the learning parameters being the corresponding degrees of rotation. Note that the top and bottom $RZ$, $RY$, and $RX$ gates labeled $p_{6}-p_{8}$ share the same parameters between pairs. Therefore, even though there are $18$ rotation gates, there are only $15$ parameters in total. 

With the encoder's unitary defined, the next step is the method of application. The unitary is applied sequentially to every pair of qubits in the circuit, shown in figures \ref{fig:encoder}b and \ref{fig:encoder}c. Note the difference between figures \ref{fig:encoder}b and \ref{fig:encoder}c is simply that figure \ref{fig:encoder}b is the encoder for the node input, which has an initial state vector length of $8$, while figure \ref{fig:encoder}c is the encoder for the edge input, which has an initial state vector length of $4$. Thus, figure \ref{fig:encoder}b requires $3$ qubits to amplitude encode the data, while figure \ref{fig:encoder}c needs $2$ qubits. When the encoder unitary is applied to different qubits, the parameters are kept the same, meaning that regardless of how many qubit pairs it is applied to the number and values of parameters is constant.

\begin{figure*}[!th]
\includegraphics[width=\textwidth,height=\textheight,keepaspectratio]{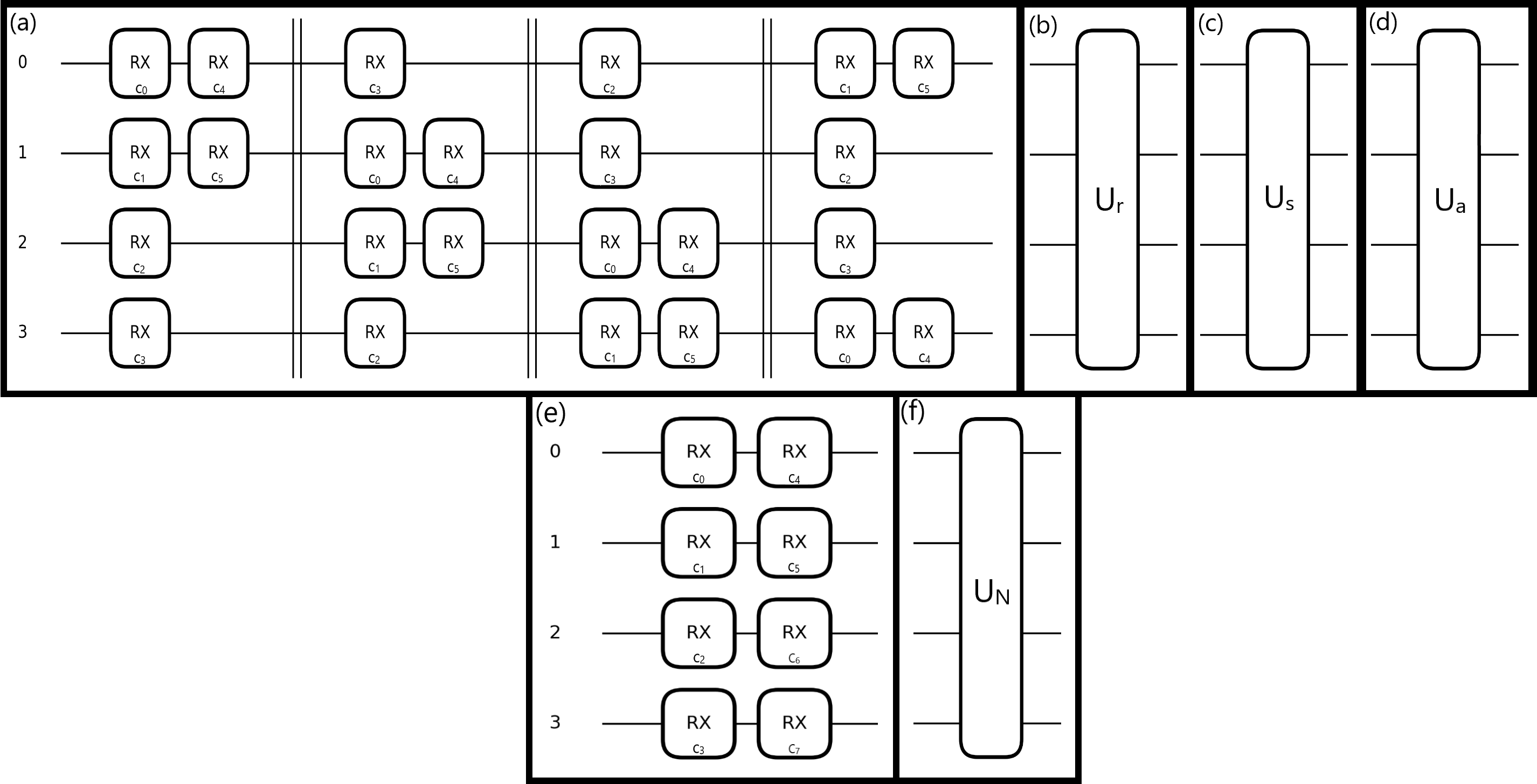}
\caption{\textbf{Implementable QGNN:} (a) Series of $RX$ gates implemented to realized values of $E_{r}$, $E_{s}$, and $E_{a}$}. (b) $RX$ unitary representation of $E_{r}$. (c) $RX$ unitary representation of $E_{s}$. (d) $RX$ unitary representation of $E_{a}$. (e) Series of $RX$ gates implemented to realized $N$. (f) $RX$ unitary representation of $N$.
\label{RX}
\end{figure*}

Concerning the decoder, it is the QCNN used by Cong et al., which is equivalent to the reverse MERA \cite{Cong}. Thus, it pools the data into a select number of qubits. For this project, the data is pooled into the final two qubits. The QCNN is realized via the repeated application of the two qubit unitary shown in figure \ref{fig:processor/decoder}c \cite{QCNN}. This unitary consists of the application of $RX$, $RY$, $RZ$, and CNOT gates, with the degrees of rotation being learned parameters. The total number of parameters is $6$. The decoder unitary is applied to pairs of qubits, with the information in the top qubit being pooled into the bottom qubit. For example, here information from qubits $0$ and $1$ is pooled into qubits $2$ and $3$, respectively as seen in figure \ref{fig:processor/decoder}d. Similar to the encoder unitary, the decoder unitary's parameters are the same in each application to qubit pairs, prior to backpropagation.

The processor is based on the design presented by Hubregtsen et. al \cite{Hubregtsen}, shown in figure \ref{fig:processor/decoder}a. It contains $16$ gates, but only $8$ parameters. It consists of two columns of $RY$ gates, each followed by cascading CNOT gates. Though the encoder and decoder are utilized only once during the entire circuit, the processor is repeatedly applied based on the desired number of message passing steps. As described in section II (also see figure \ref{fig:full_PQC_and_ALGO}), the algorithm requires the processor being applied a minimum of five times: three for edge processing and two for node processing. As is an inherent trait of GNN's, each repetition of the processor corresponds to a node's message being passed an additional node away. For example, having three repeated instances of the processor in the algorithm corresponds to a node ``knowing" about its neighbors up to three edges away \cite{Hamilton}. However, this is in the classical sense, where the processor is a single multilayer mprceptron (MLP). In the context of the quantum algorithm used in this study, a single complete run of the interaction network can be considered a single use of the processor. This is why we treat the overall edge and node processors as decomposing into their corresponding processors, as shown in equations \ref{S_E} and \ref{S_N}. Thus, another more quantitative way to consider this situation is that each $5$ uses of the processor unitary corresponds to the completion of a single message pass. Figure \ref{fig:full_PQC_and_ALGO} gives a more intutional sense of this, showing a complete run of the algorithm with the incorporation of the PQCs, containing only a single step of message passing. This figure will be explained in more detail in the following subsection. Note, table \ref{QRAM_Table} lists the number of parameters and gates found in each QGNN model given $P$ amount of processors in the learning model. As mentioned in the figure, the Implementable QGNN does not have an obvious scaling relationship with qubit count. For parameters, increasing the qubit count means expanding either the node or edge sections of the circuit or both; however, this is subjective, and depends upon the experimenter's choice. Likewise, for gate count, the implementation of the $RX$ gates would shift with increased qubit count, such being that the original series of $RX$ gates was explicitly chosen to efficiently distribute information over $4$ qubits given $6$ parameters. If either of these values changed, the application of $RX$ gates would need to be adjusted, with the manner of adjustment being arbitrary and, thus, based on the experimenter's choice. For example, given a case where there are $6$ parameters to apply and a quantum circuit with a node section containing $6$ qubits, the unitaries $U_{r}$, $U_{s}$, and $U_{a}$ of that circuit would only need a single column of $RX$ gates to effectively propagate information. This is not true for other qubit amounts. Furthermore, with increasingly large numbers of qubits, it is unclear what form the $RX$ unitaries would take once the qubit count has surpassed the $RX$ gate count of each unitary, i.e., it is arbitrary how to apply a vector of length $4$ to a quantum circuit with $7$ qubits.
\begin{table}
\includegraphics[width=0.5\textwidth]{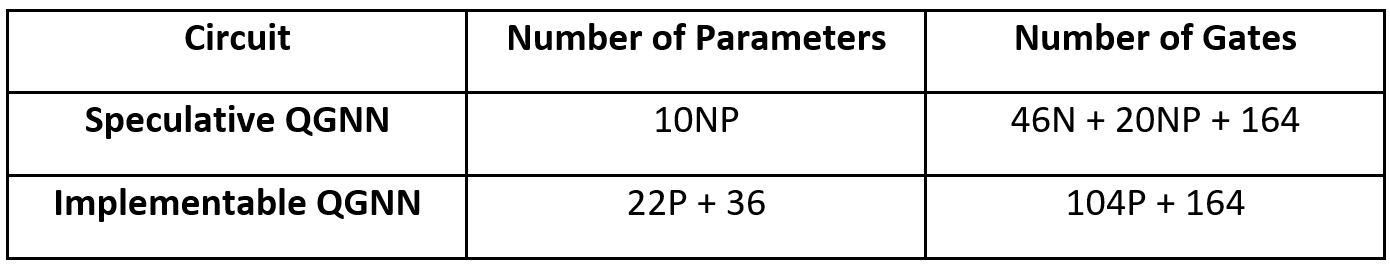}
\caption{The number of parameters and gates in each QGNN given $N$ number of qubits and the application of $P$ amount of processors. The parameters and gates for the Implementable QGNN have no obvious relationship with increased qubit count, thus this variable was left out for this circuit. Note, the gates required for the amplitude embedding were left out of these counts.}
\label{QRAM_Table}
\end{table}

\subsection{Speculative QGNN Integration}

Referring to figure \ref{fig:full_PQC_and_ALGO}, our implemented method relies on only four qubits, with the output of each sub-circuit, i.e. node encoder, $\phi_{p1}$, etc., saved and input into the next sub-circuit, as designated by the algorithm, with the corresponding rotation parameters likewise saved. The algorithm relies on numerous steps of matrix multiplication to combine information for quantifying origin, target, and magnitude of effects. This is easily done classically, where the output at each sub-circuit can be stored until the entire corresponding matrix is formed. However, doing so in a quantum circuit context would either require directly implementing superposition amplitudes or utilizing additional qubits to store the information. In the case of the latter, with multiple runs of each sub-circuit, this would lead to a quickly increasing number of qubits. This would be highly impractical, if not impossible, to realize on quantum hardware and the quantum simulation software utilized in this project, i.e. Pennylane \cite{penny_intro}. Thus, the former is necessary, even though it forfeits implementability as explained in the Introduction section. However, a QGNN that is fully implementable is realized in future sections of this paper, though it requires deviating from the prior mentioned method of propagating information via matrix multiplication.

\begin{figure*}[!th]
\includegraphics[width=\textwidth,height=\textheight,keepaspectratio]{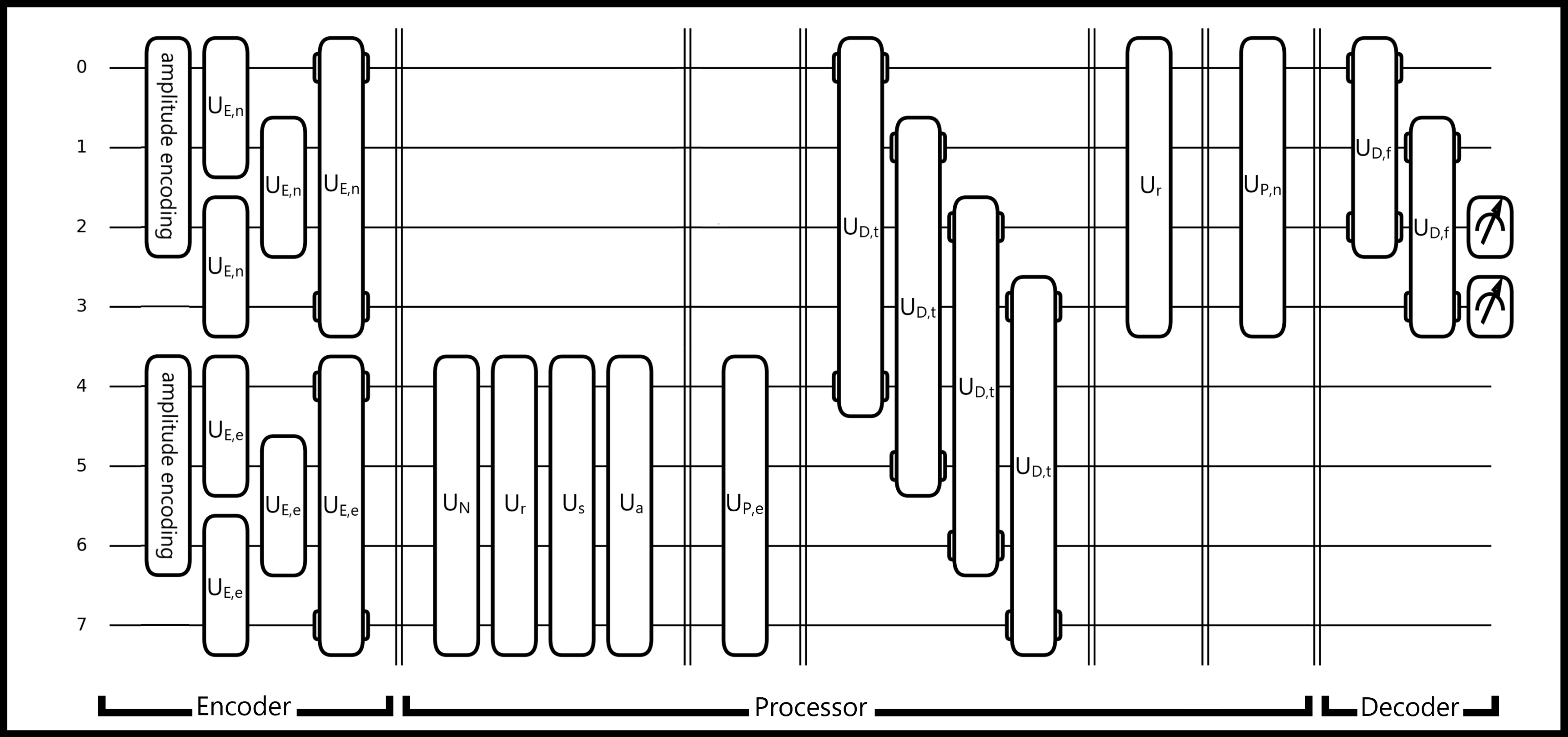}
\caption{\textbf{Implementable QGNN:} Entire Implementable QGNN quantum circuit}
\centering
\label{full-yes-QRAM}
\end{figure*}

Figure \ref{fig:full_PQC_and_ALGO} begins on the left with two encoders. The top is the node encoder and the bottom is the edge encoder. Each encoder expands the data from its initial vector length ($8$ for node input and $4$ for edge input) into a vector space of length $16$. At this point, the interaction network is implemented. The expanded versions of $N$ and $E_{a}$, as seen at the intersection of the encoders and processors in the figure, correspond to equation \ref{G}, i.e. the start of the algorithm. These superposition amplitudes are then saved, with $N$ directly plugged into the next step of the algorithm: the series of matrix multiplications and circuit applications of the edge processors as described in equations \ref{A_n}-\ref{phi_S_e}. This overall region of matrix multiplications and learning functions is the edge processor as indicated in the figure. The expanded $E_{a}$ is included in the final matrix multiplication of the edge encoder, $A_{3}$, whose output is then applied to $\phi_{e3}$. The node processor, as seen in the figure, first begins with the expanded version of $N$ being applied to $\phi_{p1}$, whose output $P_{1}'$ is then multiplied by $A$, the output of $\phi_{e3}$, and the original expanded $N$, to form $P_{2}$. The final step in the node processor is to then apply $P_{2}$ to $\phi_{n2}$, producing the updated node state $P$ (see equations \ref{P_1}$-$\ref{phi_S_n}). $P$ is then applied to the decoder, which outputs on qubits $2$ and $3$ the predicted new vertical and horizontal accelerations of the particles.

\subsection{Implementable QGNN Integration}
Examining the Implementable QGNN in figure \ref{full-yes-QRAM}, the quantum circuit is broken up into two parts: qubits $0-3$ represent the node states and qubits $4-7$ represent the edge states. It is important to first understand the implementation of the matrices $N$, $E_{r}$, $E_{s}$, and $E_{a}$ as described in equation \ref{G}. Here, the algorithm requires that the transposes of $E_{r}$, $E_{s}$, and $E_{a}$ be utilized, which is a consequence of the original shapes of these matrices. $E_{r}$ and $E_{s}$ are of size $N_{n} \times N_{e}$, and $E_{a}$ is of size $N_{l} \times N_{e}$, which, for this project means that each $E$ matrix is of shape $3 \times N_{e}$. The extra dimension of padding that $N_{l}$ had for the Speculative QGNN is not used here. However, the dimension of $N_{e}$ is variable because it describes the number of edges per given time step. Additionally, $N$ is of fixed shape $N_{l} \times N_{n}$, i.e. $8 \times 3$ for this project. Thus, for compatibility and consistency with applying the matrices in unison to the quantum circuit, the transposes of the $E$ matrices were required. In short, each applied matrix has a column size of $3$, meaning the quantum circuit is run a total of $3$ times per time step, i.e. per set of $N$, $\bar{E_{r}}$, $\bar{E_{s}}$, and $\bar{E_{a}}$, with the variability of dimension $N_{e}$ absorbed by the rotation gates (explained below).

In the Speculative QGNN, adhering to the classical learning model, information is propagated via matrix multiplication of the matrices $N$, $E_{r}$, $E_{s}$, and $E_{a}$. Here however, these matrices are applied directly to the quantum circuit via the rotation parameters of rotation gates. Figure \ref{RX}a depicts a cascading series of $RX$ gates; this entire cascade is treated as a unitary and is applied for each of the $\bar{E}$ matrices. The choice of $RX$ gate was arbitrary, though the application was purposefully kept uniform for equivalent incorporation of data. For a given $RX$ unitary, the rotation values of its $RX$ gates are determined by the row values of the given column in use. The $RX$ unitaries for $\bar{E_{r}}$, $\bar{E_{s}}$, and $\bar{E_{a}}$ are represented, respectively, by the unitaries in figure \ref{RX}b, figure \ref{RX}c, and figure \ref{RX}d. Matrix $N$, with row vector length $8$, is of greater vector length than $\bar{E_{r}}$, $\bar{E_{s}}$, and $\bar{E_{a}}$, whose row vector lengths have a possible maximum value of $6$. Thus, implementation of matrix $N$ in the quantum circuit requires two additional $RX$ gates, but only a single application for efficient information distribution amongst the qubits, as shown in figure \ref{RX}e, and its unitary representation shown in \ref{RX}f. The row vector length of the $\bar{E}$ matrices, $N_{e}$, is variable. This variability is due to the calculation of edges per time step as achieved via a nearest neighbor algorithm applied to the particles in the simulation within a certain connectivity radius. For particles within the connectivity radius of each other, an edge will be generated between them. Regardless of connections, i.e. edges, with other particles, each particle is given a self-edge. In the context of this project's simulation, there are three particles contained within a box. If no particles are within the given interaction radius of one another, then there will only be $3$ edges in that time step, each being a self-edge. However, if every particle is within the given interaction radius of every other particle, then there will be $6$ edges, three self-edges and three edges between particles. The range of possible integer number of edges is bound by these minimum and maximum values. Hence, the dimension of $N_{e}$ is bound between $3$ and $6$. Returning to the $RX$ gates, this means that the number of rotation parameters can change between each time step. To account for this, the $RX$ gate's rotation parameter assumes a value of zero when one is not provided. For example, if there exists only four edges, then the fourth and fifth $RX$ gates, $c_{4}$ and $c_{5}$ of the $E$ matrices, would be given a rotation parameter of zero.

The full application of the Implementable QGNN is shown in figure \ref{full-yes-QRAM}. In particular, this figure demonstrates a one processor implementation of the QGNN, meaning there is only one step of message passing between nodes using this circuit. For additional steps of message passing, as in the prior GNN implementations, simply add copies of the processor following the original. Each $n$ additional copy is equal to one additional run of a message passing algorithm, which corresponds to nodes learning about neighboring nodes $n$ additional steps away. Note, these copies each have their own unique parameters. The algorithm begins by encoding the node state and edge state inputs into qubits $0-2$ and $4-6$, respectively. Immediately after this, the node and edge encoder unitaries, $U_{E,n}$ and $U_{E,e}$, respectively, are applied to their corresponding sections, which are the same unitaries described by figure \ref{fig:encoder}a. Following this, the $RX$ unitaries for $N$, $\bar{E_{r}}$, $\bar{E_{s}}$, $\bar{E_{a}}$, respectively, $U_{N}$, $U_{r}$, $U_{s}$, and $U_{a}$, are applied to edge section qubits. The edge section processor, $U_{P,e}$, which is identical to that described in figure \ref{fig:processor/decoder}a, is then applied to the same section. The application of the $RX$ unitaries and processor is analog to equations \ref{A_n}-\ref{phi_S_e} of the Speculative QGNN learning model. As previously mentioned, a consequence of implementing all parts of the algorithm in a single circuit is that an extra section of decoder unitaries is required. They are used to transfer information from the edge section qubits to the node sections qubits. This requirement is inherent from the node feature state update being based upon the edge feature state update, the crux of the GNN algorithm. This transfer is done via application of the decoder unitary, represented by $U_{D,t}$; $t$ for transition from edge to node. This decoder unitary has a set of parameters different from that of the last decoder unitary, $U_{D,f}$; $f$ for final application of decoder unitary. The next unitary  is a reapplication of $U_{r}$; however, this time it is applied to the node section qubits. The use of $U_{r}$ here is analogous to equation \ref{A_BAR} of the Speculative QGNN learning model. The final unitary of the overall processor is the node section processor, $U_{P,n}$. This unitary has unique parameters from that of the edge section processor. Likewise, the application of this unitary is analogous to equations \ref{phi_n}-\ref{phi_S_n} for the Speculative QGNN. Ending the entire circuit, the decoder unitary $U_{D,f}$ is applied, and a measurement is made on qubits $2$ and $3$. The expectation value of these measurements are the predicted $x$ and $y$ direction accelerations of each particle. For a single time step, there are three cycles of this complete circuit, meaning the final output is a $2 \times 3$ matrix where the rows are the $x$ and $y$ accelerations, and the columns correspond to particular particles.

\section{Results}
\subsection{Overview}

\begin{figure*}[!th]
\includegraphics[width=\textwidth,height=\textheight,keepaspectratio]{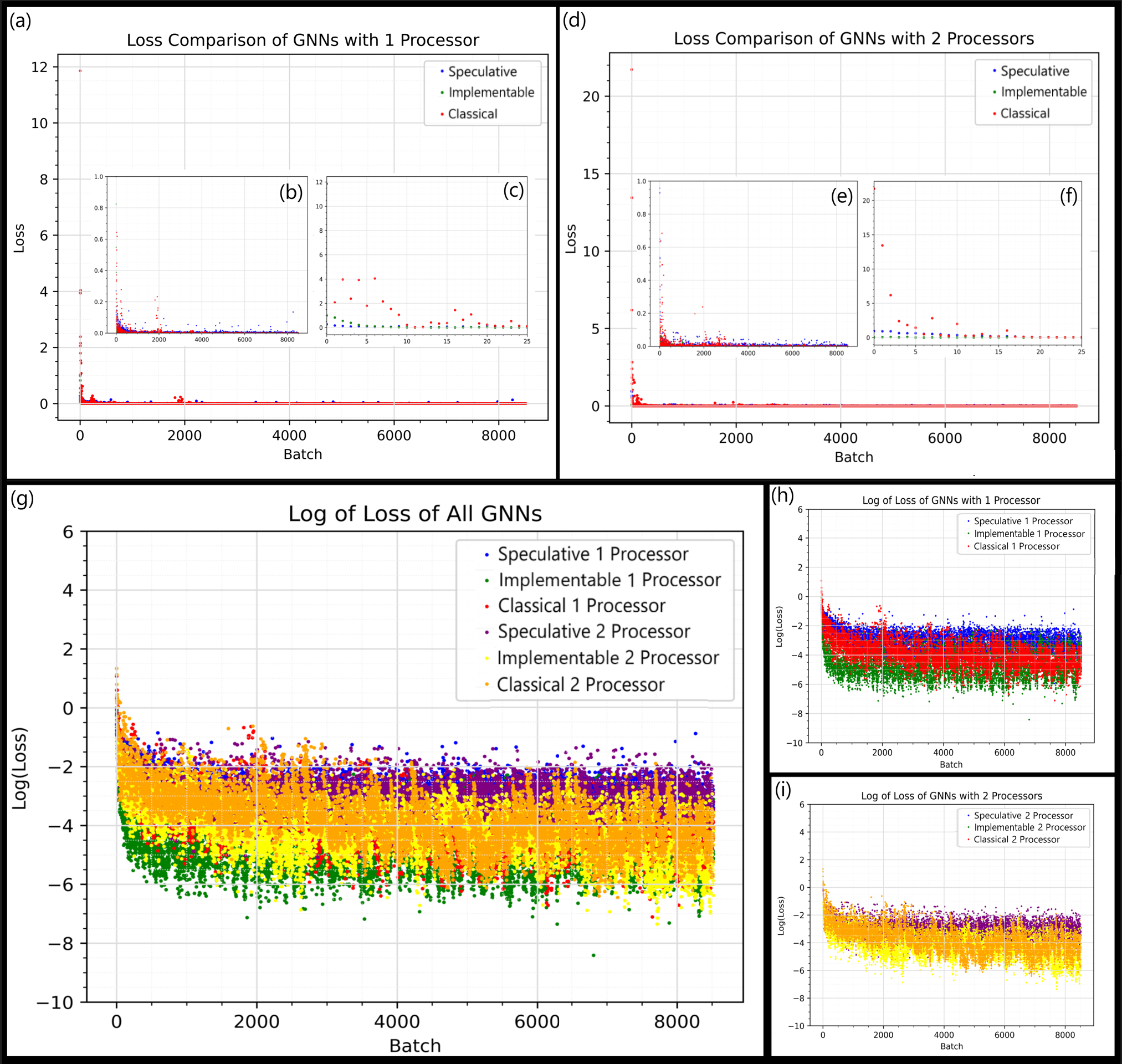}
\caption{(a) Graph of loss value over duration of training for GNNs with $1$ processor. (b) Version of (a) zoomed in on y-axis. (c) Version of (a) zoomed in on x-axis. (d) Graph of loss value per batch over duration of training for GNNs with $2$ processors. (e) Version of (d) zoomed in on y-axis. (f) Version of (d) zoomed in on x-axis. (g) Log of loss for all GNNs over duration of training. (h) Log of loss for $1$ processor GNNs over duration of training. (i) Log of loss for $2$ processors GNNs over duration of training.}
\centering
\label{Quad_Loss}
\end{figure*}

There were two groups of GNN models trained. The first consisted of GNNs with one processor, and the second consisted of GNNs with two processors. As each model progressed through its training, the calculated loss value following the application of each additional batch is shown in figure \ref{Quad_Loss}. In particular, figures \ref{Quad_Loss}a-c and \ref{Quad_Loss}d-f each show the loss value, whereas figures \ref{Quad_Loss}g-i show the common logarithm of the loss. The loss was calculated via mean square error (MSE) of the predicted acceleration of a given particle compared with its ground-truth acceleration. The ground-truth acceleration was obtained via a particle simulator found in the open source software Taichi Lang \cite{Taichi}. The predicted position of each particle was obtained via a Euler integrator that calculates the next position from the current acceleration, as implemented by Sanchez-Gonzalez et al. \cite{Sanchez-Gonzalez}. Likewise, based on Sanchez-Gonzalez et al., the optimizer implemented was Adam, featuring a learning rate of $0.01$ \cite{Sanchez-Gonzalez,Diederik}. A batch size of $4$ was used, each of the four data points being a time step, with the respective loss value of each time step being averaged together to calculate the entire batch's loss value. The data points are randomized prior to each epoch. Note, the maximum number of processors used in this project was two. This was chosen with the amount of particles in mind; in particular, the simulation consisted of three particles interacting under the influence of gravity. The graphs generated from this situation contained nodes that, at a maximum, had two-step neighbors, i.e., neighbors that are two edges away. Each processor corresponds to a single step; thus, the maximum number of processors needed was two.

\begin{figure*}[!th]
\includegraphics[width=\textwidth,height=\textheight,keepaspectratio]{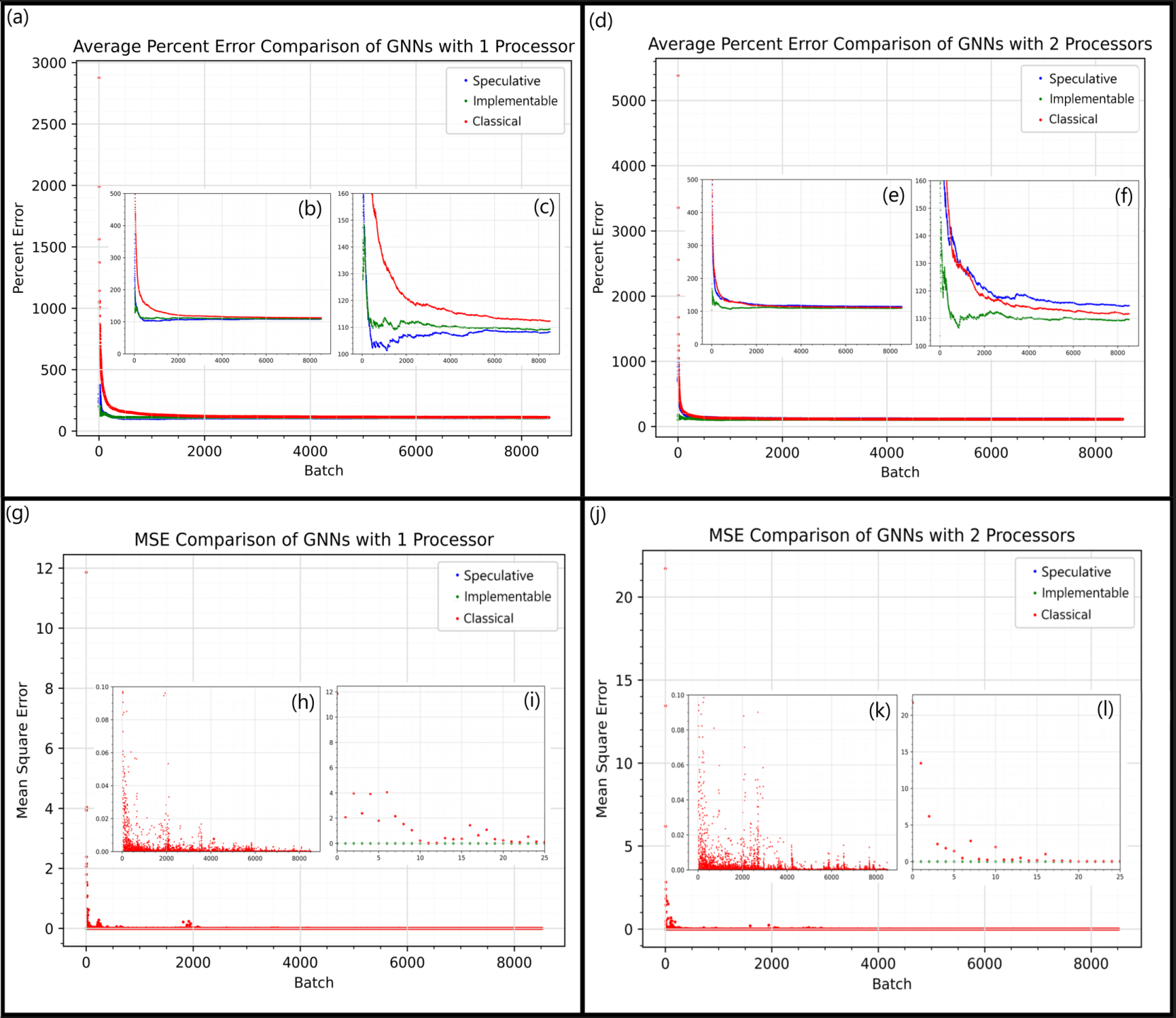}
\caption{(a) Graph of average percent error in position prediction over duration of training for GNNs with $1$ processor. (b) Version of (a) moderately zoomed in on y-axis. (c) Version of (a) highly zoomed in on y-axis. (d) Graph of average percent error in position prediction over duration of training for GNNs with $2$ processors. (e) Version of (d) moderately zoomed in on y-axis. (f) Version of (d) highly zoomed in on y-axis. (g) Graph of MSE error in position prediction over duration of training for GNNs with $1$ processor. (h) Version of (g) zoomed in on y-axis. (i) Version of (g) zoomed in on x-axis. (j) Graph of MSE error in position prediction over duration of training for GNNs with $2$ processors. (k) Version of (j) zoomed in on y-axis. (l) Version of (j) zoomed in on x-axis. }
\centering
\label{Quad_MSE}
\end{figure*}

\subsection{Learning Efficiency}
Figure \ref{Quad_Loss}a shows the entire loss value trajectory as training progressed for the GNNs with a single processor. As evident in this figure, there exists considerable overlap between the loss values for the classical GNN (CGNN), Implementable QGNN, and Speculative QGNN. Figures \ref{Quad_Loss}b and \ref{Quad_Loss}c show zoomed in views. The prior zooms into the y-axis, looking between a range of $[0,1]$, while the latter zooms into the x-axis, looking between a range of $[0,25]$. Figure \ref{Quad_Loss}b further demonstrates the considerable overlap between loss values, and only in the figure \ref{Quad_Loss}c does the difference become clear. The CGNN starts with the highest loss value, the Implementable QGNN starts in the middle, and the Speculative QGNN starts with the lowest. Additionally, the QGNNs approach a near zero loss value approximately $5$ batches prior to the classical. Regardless, each GNN approaches this near zero amount within $10$ batches. It is difficult to compare the learning efficiencies of the $1$ processor GNNs in this direct manner; thus, examination of their logarithmic plots was necessary, as shown in figure \ref{Quad_Loss}h. It is evident that the Implementable QGNN reaches and maintains the lowest loss values, followed by the CGNN, and last by the Speculative QGNN. Considering each GNN approaches a near zero value within the first one percent of applied batches, using the logarithmic results as criteria for comparison of learning efficiencies is appropriate. Thus, the Implementable QGNN is most efficient, the CGNN is second, and the Speculative QGNN is least. However, the de facto results show the learning efficiency of each $1$ processor GNN model is highly similar, with a notable degree of overlap existing between loss values throughout training. Overall, each model is highly efficient in reducing the loss throughout training.

Figure \ref{Quad_Loss}d shows the entire loss value trajectory as training progresses for the $2$ processor GNNs. Additionally, figures \ref{Quad_Loss}e and \ref{Quad_Loss}f show zoomed in views of the y-axis and x-axis as described in the $1$ processor GNNs case. Likewise, analysis of these figures, in addition to the logarithmic plot of the $2$ processor GNNs loss values, as seen in figure \ref{Quad_Loss}i, will result in the same conclusions made for the $1$ processor GNNs case. Specifically, in the context of learning efficiency, the Implementable QGNN is most efficient, the CGNN is second, and the Speculative QGNN is least efficient. However, once again, the de facto results show the learning efficiency of each $2$ processor GNN model is highly similar, with a notable degree of overlap existing between loss values throughout training. Overall, each model is efficient in reducing the MSE loss throughout training. These conclusions are based on the same observations as found in the $1$ processor case.

\begin{figure*}[!th]
\includegraphics[width=\textwidth,height=\textheight,keepaspectratio]{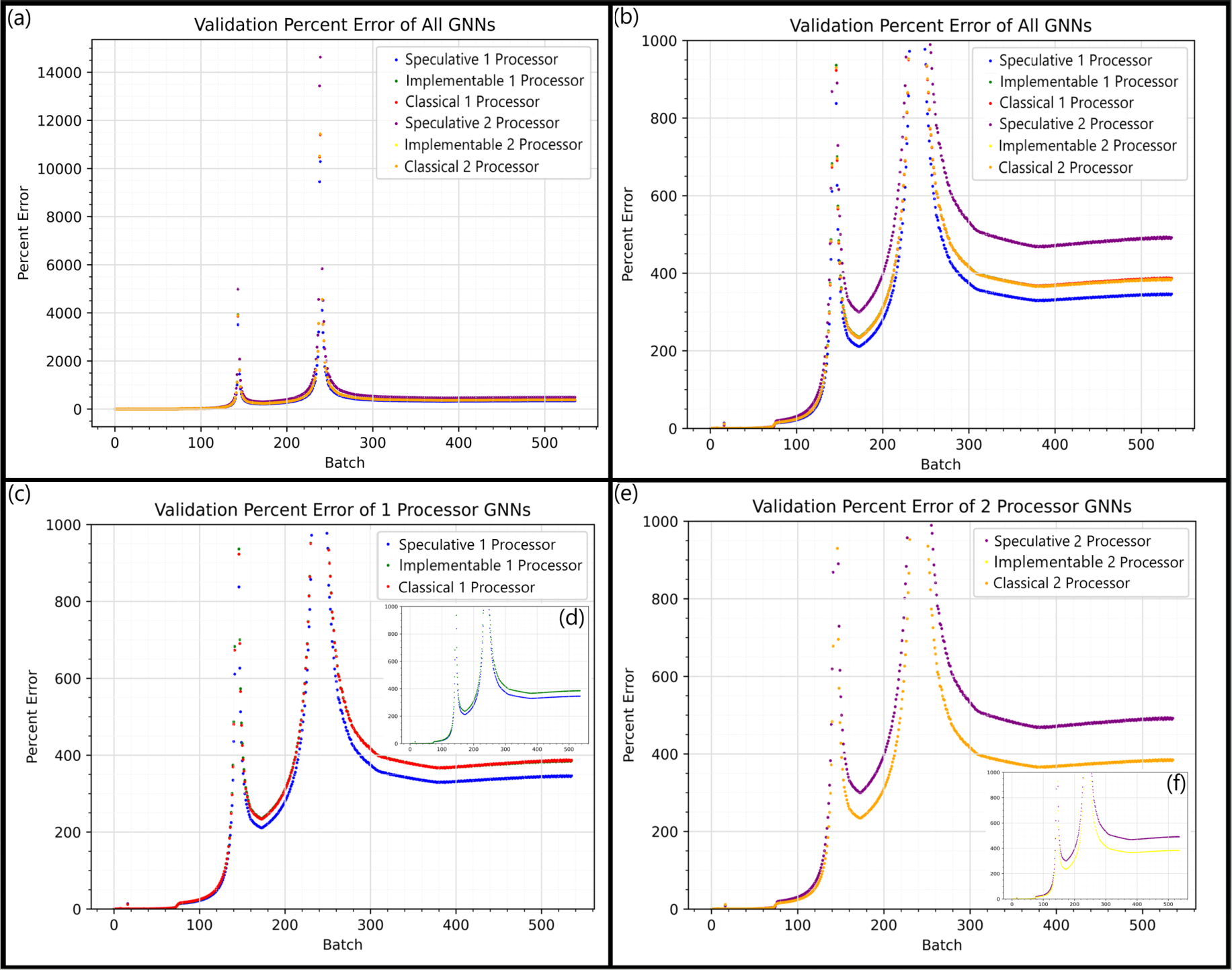}
\caption{Note: The high percent errors in the above graphs are not indicative of fruitless performances by each model; see Performance and Hyperparmeters section for full explanation.(a) Percent error using validation data set with all GNNs. (b) Zoomed in view of (a). (c) Percent error using validation data set with $1$ Processor GNNs. (d) Same as (c) except with the CGNN omitted to show its overlap with the Implementable QGNN. (e) Percent error using validation data set with $2$ Processor GNNs. (f) Same as (e) except with the CGNN omitted to show its overlap with the Implementable QGNN.} 
\centering
\label{fig:validation}
\end{figure*}

Figure \ref{Quad_Loss}g shows the logarithmic plot of loss values for each GNN in both the $1$ processor and $2$ processor cases. As a general trend, it appears that the Implementable QGNNs have the highest learning efficiencies, the classical GNNs both have the middle, and the Speculative QGNNs both have the lowest. However, comparing model pairs, the $1$ processor GNNs are more efficient than their $2$ processor counterparts. This suggests that the graphs generated in the ground-truth three-particle simulation consist mainly of one-step graphs, i.e. graphs containing nodes whose neighbor nodes are only a single edge away. This would explain the difference in efficiency because $2$-processor GNNs used on graphs containing nodes with only $1$-step neighbors would be redundant. In particular, consider that each additional processor corresponds to a particular node learning about a node an additional step away, i.e. message passing. In this situation, the second message pass would be redundant because there would be no $2$-step neighbor to learn about. Furthermore, the second message pass may cause an over mixing of the node states, decreasing the distinct features of each node, reducing useful information in the system.

\subsection{Accuracy}
Though it is positive to observe that each GNN has a high learning efficiency, it is equally important to observe the accuracy in the model predictions. To measure this, two methods were used; the first was taking the MSE of the predicted and target next position values of each particle, and the second was taking the percent error of these same values. The prior is based on Sanchez-Gonzalez et al. work, in which they utilized the same means of accuracy measurement \cite{Sanchez-Gonzalez}. The latter is based on the observation that the positions, predicted and actual, of each particle consist of considerably small numbers, a majority of the time being between $[-1,1]$. Thus, MSE measurements will already be a near zero value, meaning that the MSE method of estimating accuracy is impractical for visual comparison in the context of this study's results. Regardless, some useful information can still be obtained from viewing the MSE plot of each GNN. 

\begin{figure*}[!th]
\includegraphics[width=\textwidth,height=\textheight,keepaspectratio]{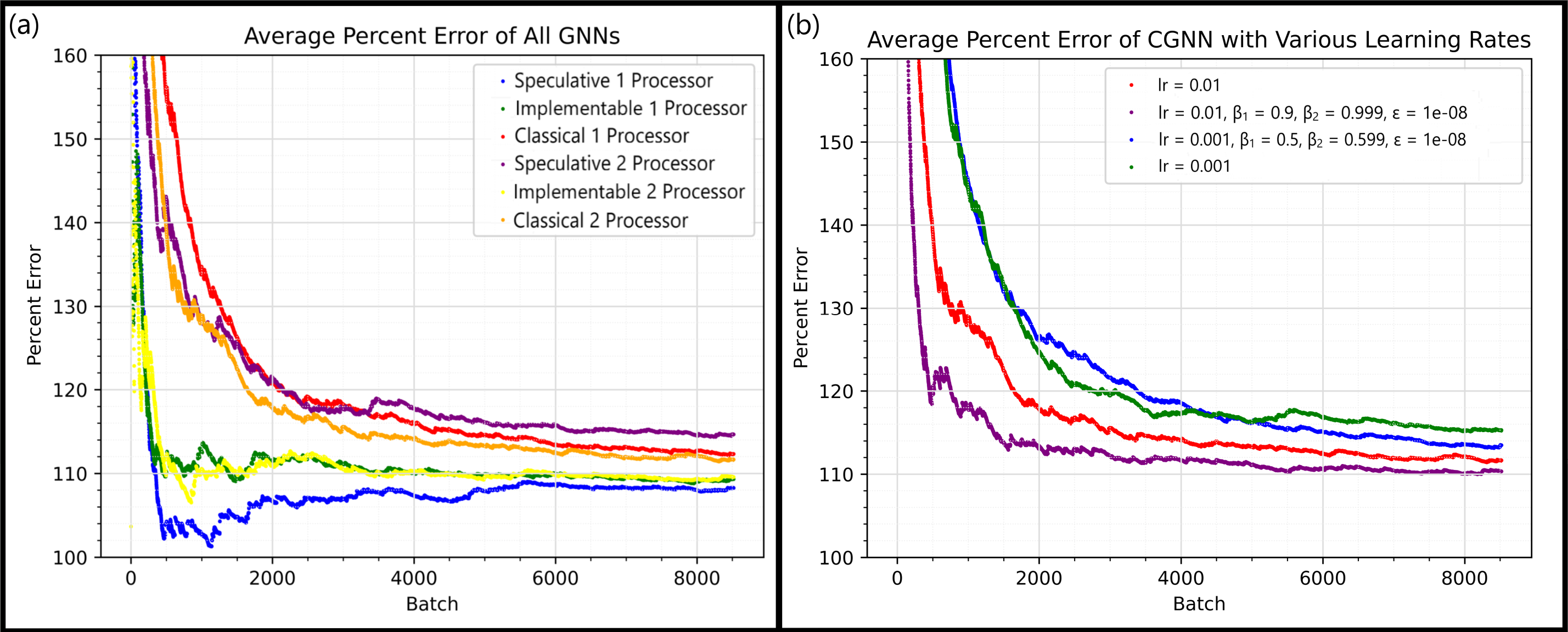}
\caption{(a) Average percent error in position prediction over duration of training for all GNNs. (b) Average percent error in position prediction over duration of training for CGNN with various learning rates. Proceeding down the legend, the first learning rate is the original, i.e., the learning rate implemented throughout this study, the second and third learning rates are variations that adjust throughout training, and the fourth learning rate is a variation that is simply smaller than the original. Here, ``lr" stands for learning rate and $\beta_{1}$, $\beta_{2}$, and $\epsilon$ are the relevant variables to the Adam algorithm they are implementing.}
\centering
\label{Double_Results}
\end{figure*}

Figures \ref{Quad_MSE}g-i and \ref{Quad_MSE}j-l show the MSE plots for the $1$ processor and $2$ processor GNNs, respectively. Figures \ref{Quad_MSE}h and \ref{Quad_MSE}k zoom into the y-axis, looking between a range of $[0,0.10]$, while figures \ref{Quad_MSE}i and \ref{Quad_MSE}l zoom into the x-axis, looking between a range of $[0,25]$. Similar to the learning rates for both $1$ processor and $2$ processor cases, the MSE value rapidly decreases to near zero values. Likewise, the CGNN begins with the highest MSE values, and realigns with the Implementable QGNN and Speculative QGNN at approximately $10$ batches. However, here the Implementable QGNN and Speculative QGNN immediately begin with near zero MSE values. Overall, the MSE values of the $1$ processor GNNs overlap considerably, as do the MSE values of the $2$ processor GNNs. Note that as a general trend the $2$ processor GNNs have a higher MSE value throughout training as compared to the $1$ processor GNNs. Furthermore, they do take more batches to reach the same near zero MSE values as already reached by the $1$ processor GNNs.

Figures \ref{Quad_MSE}a-c and \ref{Quad_MSE}d-f show the percent error plots for the $1$ processor and $2$ processor GNNs, respectively. Figures \ref{Quad_MSE}b and \ref{Quad_MSE}e moderately zoom into the y-axis, looking between a range of $[0,500]$, while figures \ref{Quad_MSE}c and \ref{Quad_MSE}f highly zoom into the y-axis, looking between a range of $[100,160]$. Note, the graphs of percent error are averages, with the average percent error calculated, i.e. updated, at every batch, and the resulting average percent error plotted. Examining the $1$ processor GNNs, it is immediately clear that they deviate from the results of the learning efficiency comparisons. In particular, here, the Speculative QGNN has the highest accuracy throughout training, followed by the Implementable QGNN, and last by the CGNN. This is in direct contrast to the Implementable QGNN having the greatest learning efficiency, followed by the CGNN, and last by the Speculative QGNN. However, examining the $2$ processor GNNs, they do not have this deviation but instead follow the pattern established by their learning efficiencies. This difference is a possible consequence of the occasional redundancy of the $2$ processor GNNS in the case of this three-particle simulation, as prior described. Likewise, this is also a possible consequence of the increase in parameters with the inclusion of an additional processor, which does not increase the parameter count equally in each GNN model. Regardless, as shown in figure \ref{Double_Results}a, in both the $1$ processor and $2$ processor cases, the percent error of the Implementable QGNN, Speculative QGNN, and CGNN all decrease at comparable rates, leveling off in close proximity approximately at $110\%$ error for the $1$ processor GNNs and at $112\%$ error for the $2$ processor GNNs. Note, the high degree of inaccuracy shown in these measurements must be considered in the context of all other results. Thus, these measurements do not indicate that the performances of these models are fruitless; for a full analysis, see the Performance and Hyperparmeters section below.

It is worth comparing percent error of all the GNNs together, as shown in figure \ref{Double_Results}a. Here, there is no particular pattern to the accuracy rankings. The Speculative QGNN with $1$ processor performs the best, while its $2$ processor counterpart performs the worst. The Implementable QGNNs perform second and third best, with both the $1$ and $2$ processor cases performing approximately the same. Likewise, the CGNNs perform forth and fifth best, with the $2$ processor case performing slightly better overall then the $1$ processor case. Regardless, these differences ultimately are rather minute, with each GNN having a difference in percent error accuracy within at most approximately $10\%$ of each other. 

\begin{figure*}[!th]
\includegraphics[width=\textwidth,height=\textheight,keepaspectratio]{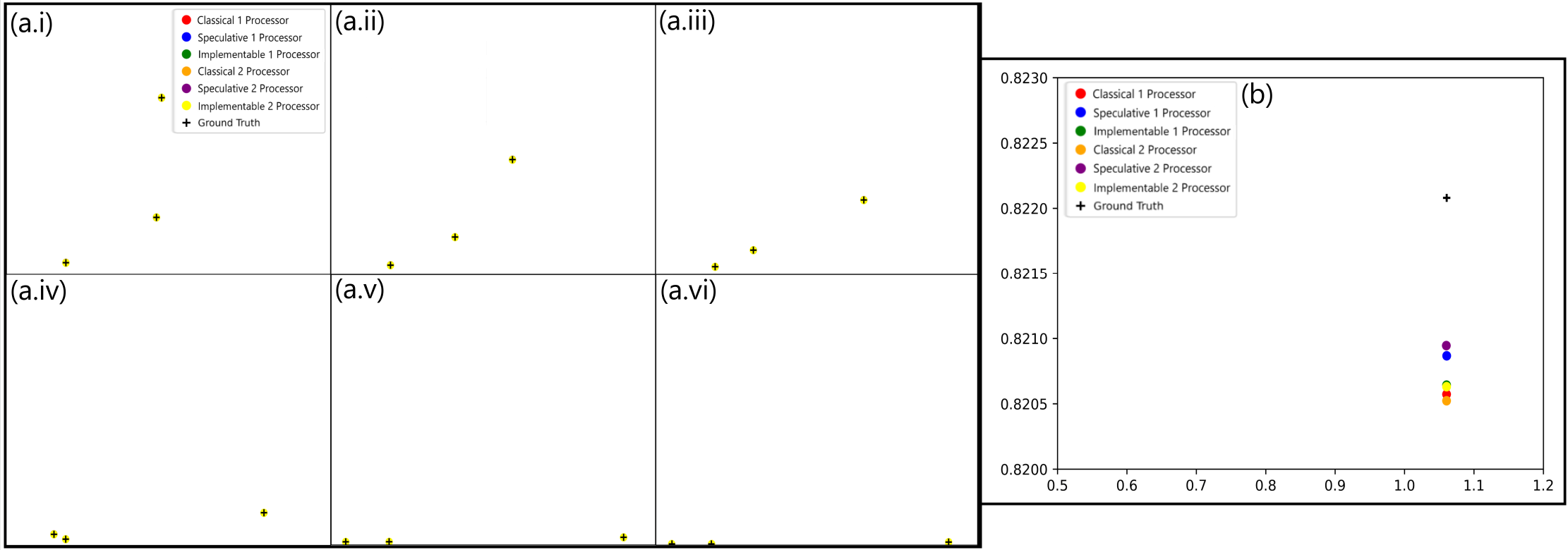}
\caption{(a.i)-(a.vi): A select sample of predicted positions given by each GNN model using the validation data set. The time progression is incremental, beginning at (a.i) and ending at (a.vi). The particles begin off the ground, and their progression as they fall and collide can be approximately observed. (b) Zoomed in view of the rightmost particle in graph (a.iii).}
\centering
\label{time_prog}
\end{figure*}

We find a similar situation testing the trained models on the validation data set, which is approximately $30\%$ the size of the training data set. The time progression of sampled position predictions made by each model using this data set can be seen in figure \ref{time_prog}a and the results can be seen in figure \ref{fig:validation}. In particular, figure \ref{fig:validation}a shows the percent error in predictions for all models while running the validation data set. The accuracy rankings given by the training results are comparable to the outcomes here. Figure \ref{fig:validation}b demonstrates this; zooming in on the y-axis, the $1$ processor Speculative QGNN performs the best, while its $2$ processor counterpart performs the worst. In this case however, the performance of the remaining models is nearly indistinguishable, being almost completely overlapped. For completeness, figures \ref{fig:validation}c-f show the zoomed in views of the Implementable QGNN and CGNN cases. In particular, figure \ref{fig:validation}c shows the $1$ processor GNNs case, and figure \ref{fig:validation}e shows the same except with the CGNN results omitted to prove they overlap with the results of the Implementable QGNN. Likewise, figure \ref{fig:validation}e shows the $2$ processor GNNs case, and figure \ref{fig:validation}f shows the same except with the CGNN results omitted to prove they overlap with the results of the Implementable QGNN. 

The accuracy in the context of the validation data set is notably worse based on the percent error measurements. However, this is not surprising, as the percent error accuracy was initially poor throughout training. Additionally, the overall performance of these models is less similar than their performances during training. In particular, for the final half of percent error results, the $1$ processor Speculative QGNN resides at approximately $350\%$, both cases of the Implementable QGNNs and CGNNs reside at approximately $400\%$, and the $2$ processor Speculative QGNN resides at approximately $500\%$.

\subsection{Performance and Hyperparmeters}
In determining the performance of the GNNs, it is necessary to consider their varying measurements of accuracy. In particular, their MSE measurements, combined with observing the constant overlap of particles in figure \ref{time_prog}a, would suggest that their accuracies are high. These combined observations indicate that each model is capable of following the general trend of the ground-truth. However, this must also be considered in the context of the percent error measurements and figure \ref{time_prog}b, the zoomed in view of the rightmost particle in figure \ref{time_prog}a.iii. The method of simulating particle interactions is via generating time steps with a small time increment between consecutive steps. For the ground-truth simulation this time increment was $0.0001$, meaning particle movement behaves approximately to this scale. Figure \ref{time_prog}b shows that the vertical distance between the closest particle is $0.001$ (arbitrary units). This difference is $10$ times greater than the time increment. Thus, this large difference indicates that percent error measurements are also correct, meaning each model has notable degree of inaccuracy. Considering the conclusions based on both the MSE error and percent error measurements, the GNN models are hence moderately inaccurate. To be precise, they able to approximate the general trend of particle interactions while being a non-negligible percent off.     

The moderate inaccuracy and high learning efficiency of each model suggests that they are able to quickly identify some simple features in the data, and accurately make predictions based on them, which results in the high learning efficiency. However, simultaneously there are more complex variables at work, for which the models are completely inept at determining, resulting in an overall moderate degree of inaccuracy. We found that these complexities are related to the nature of the problem: particles in a box interacting under the influence of gravity. If pushed in the x-direction, there are no forces on the particles except for collisions with boundaries or other particles. If they are falling, bouncing off another particle, or doing some combination of these actions, this is a complex behavior. This is exemplified in figure \ref{fig:validation}, where the sudden peaks in the percent error correspond to particles falling under the influence of gravity and particles colliding, whereas the remaining portions of the graphs correspond to particles rolling. This is further confirmed in the raw data, where it was observed that the x-direction values of the particles tended to be considerably more accurate than the y-direction values (see also figure \ref{time_prog}b). Additionally, this suggests that the percent error in the training plateaued around $100\%$ as an effect of the x-direction values being accurate, while the y-direction values were incorrect to a notable magnitude.

The moderate inaccuracy of the models does not condemn them as a whole. Rather rudimentary neural network structures were used throughout this project, with similarly simple learning rates. Any increasingly advanced techniques, such as dropout and decaying learning rates, were avoided. This lack in optimization of hyperparameters is likely a large contributor to the current issues with these models. This notion is further supported by figure \ref{Double_Results}b, which shows the percent error trajectories of the $2$ processor CGNN for the learning rate of $0.01$ compared to various learning rates. Proceeding down figure \ref{Double_Results}'s legend, the first two variations follow the Adam learning rate algorithm described by Kingma and Ba \cite{Diederik}, with the relevant variables given in the legend accounting for the difference in performance. The third variation is simply a decrease in the learning rate's magnitude. As shown by the first learning rate variation, a simple adjustment of this parameter already results in an increase of accuracy. As mentioned prior, the models implemented in this study were purposely kept simple for the sake of ease and efficiency in implementation. Thus, the basic learning rate of $0.01$ was used throughout training and testing.  

As described at the beginning of this study, following proof of a quantum GNN analog, the results of the CGNN and Speculative QGNN were obtained to compare against the Implementable QGNN. It is promising that the Implementable QGNN has a greater learning efficiency than both the CGNNs and Speculative QGNNs. Likewise, during the training, the Implementable QGNN's accuracy performance appears to offer advantage over the CGNNs and the Speculative QGNN (ideal quantum-classical GNN analog) in situations containing redundancy. However, the identical performances of the CGNN and Implementable QGNN when tested on validation data suggest that this question requires further experimentation to reach any definite conclusion. Furthermore, this study was completed using a quantum circuit simulator, and thus would have to be implemented on actual quantum hardware to determine any real advantage.

\section{Conclusion}

The aim of this project was to construct quantum analogs to the classical graph neural network, as based on the work of Sanchez-Gonzalez et al. \cite{Sanchez-Gonzalez}. That goal was realized via two quantum graph neural networks, one that was speculative and one that was implementable. These QGNNs were compared alongside the CGNN in the task of particle interaction simulation. For simplicity, the case of three particles contained within a box was used to generate the training data; likewise, the most basic form of these GNNs were implemented. Two sets of GNNs were tested. The first contained a single processor, and the second contained two processors. Overall, the models proved capable of learning simple characteristics in the data, resulting in a high learning efficiency. However, they were unable to determine the more complex behaviours that were simultaneously occurring, resulting in a moderate inaccuracy in predictions. These conclusions are evident in the discrepancy between the predictions in the x and y values, in which the x value predictions tended to be more accurate because they are governed by simpler behaviors.

In addition to the successful realization of QGNNs, the results of this study suggest that the Implementable QGNN could have an advantage over CGNNs in learning efficiency and accuracy. However, further testing is required to confirm this. Furthermore, it is likely that the overall moderate inaccuracy in predictions is not wholly a fault of the models, but perhaps a result of not fine-tuning the hyperparameters. This leads to a path for a potential future study. In particular, future research should implement these models under a variety of hyperparamters, observing the consequences on learning efficiency and accuracy.

\acknowledgments{
        The views expressed are those of the authors and do
        not reflect the official guidance or position of the United
        States Government, the Department of Defense, the
        United States Air Force or the Griffiss Institute. 
        The appearance of external hyperlinks does not constitute endorsement by the United States Department of Defense of the linked websites, or the information, products, or services contained therein. The Department of Defense does not exercise any editorial, security, or other control over the information you may find at these locations.}

\section*{Data Availability}
The data that support the findings of this study are available from the corresponding author upon reasonable request.
\newpage
\nocite{*}
\bibliography{biblio}

\end{document}